\documentclass[pre,aps,eqsecnum,psfig]{revtex4}

\begin{document}

\title{Entropy production away from the equilibrium}
\author{Gregory Falkovich and Alexander Fouxon}
\address{Physics of Complex Systems, Weizmann Institute of Science,
Rehovot 76100 Israel}

%\date{\today}

\begin{abstract}For a system
moving away from equilibrium, we express the entropy production
via a two-point correlation function for any time and any distance
from equilibrium. The long-time limit gives the sum of the
Lyapunov exponents for a general dynamical system expressed via
the formula of a Green-Kubo type.
\end{abstract}
\maketitle

%\documentclass[pre,aps,eqsecnum,psfig]{revtex4}

%\begin{document}

%\title{Entropy production in random flows and
%dynamical systems} \maketitle

\section{Introduction}
We discuss here how systems go away from the equilibrium under the
action of external factors. An appropriate way to describe the
process is to study the evolution of the density $n$ which
satisfies the continuity equation
\begin{equation}
{\partial n\over\partial t}+{\rm div}\,n{\bf v}=0\
.\label{basic}\end{equation} Here $n(t, {\bf r})$ is either a
phase-space density or just plain density in space and ${\bf v}(t,
{\bf r})$ is either a velocity field defining the phase space
dynamics or the velocity field in space.  We consider a compact
phase space and assume that the total mass is conserved: $\int
n(t, {\bf r})\,d{\bf r}=1$. Equilibrium is generally characterized
by smooth measures like microcanonical distribution over the
energy surface in the phase space or just uniform density in space
when the flow is incompressible. Equilibrium state realizes the
maximum of the Gibbs entropy
\begin{eqnarray}&& S=-\int n({\bf r})\ln
n({\bf r})d{\bf r}\ \label{entropy1}
\end{eqnarray}
and is given by $n=const=1/V$. This state is not stationary if
${\bf v}(t, {\bf r})$ is compressible that is $\omega={\rm
div}\,{\bf v}\neq 0$. If the system is initially at equilibrium
then a compressible flow move it away from it.  The density field
$n({\bf r},t)$ is getting inhomogeneous and the Gibbs entropy
starts to decrease. This decrease can be interpreted as the
entropy flux from the system to the environment that provides for
compressibility of the velocity field (for a closed Hamiltonian
system $\omega=0$). Our main interest here is in the entropy
production rate which can be expressed via the velocity
divergence:
\begin{eqnarray}&&
\frac{dS}{dt} =\int d{\bf r}n({\bf r}, t)\omega({\bf r},t) =\int
\omega({\bf q}(t,{\bf x}),t)\,\frac{d{\bf x}}{V}\ ,\label{e1}
\end{eqnarray}
The main feature of our approach is the application of Lagrangian
description with trajectories ${\bf q}(t,{\bf x})$ defined by
$\dot{\bf q}={\bf v}({\bf q},t)$ and ${\bf q}(0,{\bf x})={\bf x}$.
Our main result derived below is the formula for the entropy
production rate at an arbitrary time in terms of the two-point
correlation function along the Lagrangian trajectories:
\begin{eqnarray}
&&\frac{d^2S}{dt^2} =-\int \omega({\bf x},0)\omega({\bf q}(t,{\bf
x}),t)\, \frac{d{\bf x}}{V}\equiv \langle\omega(0)\omega(t)\rangle
\label{d2S}\\
&&\frac{dS}{dt} =-\int_0^tdt \langle\omega(0)\omega(t')\rangle dt'
\label{d1s}
\end{eqnarray}
The formula holds either for time-independent ${\bf v}({\bf r})$
or for a statistically steady ${\bf v}(t, {\bf r})$. Near
equilibrium, (\ref{d2S}) is a well-known Green-Kubo formula (see,
e.g. (5.7) in \cite{Ruelle}). It is remarkable that the entropy
production is expressed via the pair correlation function not only
near equilibrium  but in any state.

Equation (\ref{d1s}) gives $dS/dt\leq0$ at short times which is
physically quite natural since the entropy has nowhere to go but
down. When the velocity field is steady (or statistically steady)
the density field may tend to a non-equilibrium steady state (see
\cite{Dorfman,Ruelle} and the references therein). Examples of
such states are Sinai-Ruelle-Bowen measures in hyperbolic
dynamical systems \cite{Sinai,BR75}. The steady state is
characterized by the limiting entropy production rate
$\lim_{t\to\infty} {\dot S}=\sum\lambda_i$. Here $\sum \lambda_i$
is the sum of Lyapunov exponents $\lambda_i$ if the latter exist.
This sum is generally non-positive \cite{Ruelle97,BFF99}. We
observe that it obeys the formula that looks exactly the same as
the Green-Kubo formula:
\begin{eqnarray}&&
\sum \lambda_i=-\int_0^{\infty} \langle\omega(0)\omega(t)\rangle
dt\leq 0.
\end{eqnarray}
If the non-equilibrium steady state density $\mu$ is smooth then
its Gibbs entropy is finite and necessarily
$\lim_{t\to\infty}{\dot S}= \sum\lambda_i=0$. It is often the case
that $\mu$ is singular, the simplest example is steady $v(r)$ in
one dimensional case where $n$ accumulates in the stagnation
points. Generally for such states ${\dot S}=\sum\lambda_i<0$ that
does not contradict stationarity since $S=-\infty$ where the
measure is singular. Constant flux of entropy from the system to
the environment is an important feature of these states.
%One can
%also introduce a coarse-grained (Boltzmann) entropy which
%asymptotically saturates; the ever-increasing difference between
%the Gibbs entropy and the Boltzmann entropy can be interpreted as
%an entropy production of a non-equilibrium system \cite{BTV}.

We see that the entropy production has the same sign both at the
beginning and at the end of the evolution. How it behaves in
between may depend on the system.  In particular, it follows from
(\ref{d2S}) that when two-point correlation function of the
velocity divergence does not change sign, the entropy production
rate ${\dot S}$ monotonically decreases from $0$ in equilibrium to
its (negative) minimum realized in the non-equilibrium steady
state with singular density $\mu$. Entropy decrease corresponds to
the increase of order in the system.

\section{General considerations}\label{sec:generalissimus}

We consider first a smooth dynamical system defined by a steady
velocity field ${\bf v}$:
\begin{eqnarray}&&
\frac{d{\bf x}}{dt}={\bf v}({\bf x}),
\end{eqnarray}
where ${\bf x}$ is the phase space coordinate of the system. For
most part we assume that the phase space is the finite volume $V$
region in $d-$dimensional Euclidean space even though
generalizations are possible. Velocity field defines the flow
${\bf q}(t, {\bf r})$ in the phase space by
\begin{eqnarray}&&
\frac{\partial {\bf q}(t, {\bf r})}{\partial t} ={\bf v}({\bf
q}(t, {\bf r})),\ \ {\bf q}(0, {\bf r})={\bf r}.
\end{eqnarray}
Let us define
$$W_{ij}(t, {\bf r})\equiv \frac{\partial q_i(t,
{\bf r})}{\partial r_j}\ .$$ An important relation between ${\bf
q}(t, {\bf r})$ and ${\bf v}({\bf r})$ holds
\begin{eqnarray}&&
{\bf v}({\bf q}(t, {\bf r}))=W(t, {\bf r}){\bf v}({\bf r})\,,
\label{ooo}\end{eqnarray} which allows, in particular, to describe
the evolution of the functions $f({\bf q}(t, {\bf r}))$ defined on
trajectories. Using (\ref{ooo}) we have for any differentiable $f$
\begin{eqnarray}&&
\frac{d}{dt}f({\bf q}(t, {\bf r}))=({\bf v}({\bf r})\cdot
\nabla_r) f({\bf q}(t, {\bf r}))=-\omega({\bf r})f({\bf q}(t, {\bf
r})) +\nabla_r\cdot [{\bf v}({\bf r})f({\bf q}(t, {\bf r}))].
\label{deriv}\end{eqnarray} We now integrate it over the space:
\begin{eqnarray}&&
\frac{d}{dt}\int \frac{d{\bf r}}{V}f({\bf q}(t, {\bf r}))=-\int
\frac{d{\bf r}}{V} \omega({\bf r})f({\bf q}(t, {\bf
r}))\equiv-\langle \omega(0) f(t)\rangle\,,\label{dEdt}\\
&&\int \frac{d{\bf r}}{V}f({\bf r})- \int \frac{d{\bf r}}{V}f({\bf
q}(t, {\bf r}) =\int_0^t \langle \omega(0)f(t')\rangle dt'\,,
\label{cor}\end{eqnarray} where we assumed that the space integral
of the last term in (\ref{deriv}) vanishes because it can be
written as an integral over the boundary. This is true in the
periodic case or in the case where the normal component of
velocity vanishes (it might be interesting to consider the cases
where boundary is important as well). We defined the correlation
function
\begin{eqnarray}&&
\langle g(0)f(t)\rangle\equiv \int \frac{d{\bf r}}{V} g({\bf
r})f({\bf q}(t, {\bf r}))\ . \label{oo}\end{eqnarray} We shall see
below that our main result (\ref{d2S}) is a particular case of
(\ref{dEdt}). Integrating the equation (\ref{cor}) over time we
have
\begin{eqnarray}&&
\int \frac{d{\bf r}}{V}f({\bf r})-\frac{1}{t}\int_0^t dt' \int
\frac{d{\bf r}}{V}f({\bf q}(t', {\bf r})= \int_0^t \langle
\omega(0)f(t')\rangle dt'-\frac{1}{t}\int_0^t t' \langle
\omega(0)f(t')\rangle dt'. \label{cor1}\end{eqnarray} Let us
consider what may happen with Equations (\ref{cor}) and
(\ref{cor1}) when $t\to\infty$. From Equation (\ref{cor}) we
observe that provided $\int_0^{\infty}\langle \omega(0)f(t)\rangle
dt$ exists there must also exist a finite limit of the spatial
average
\begin{eqnarray}&&
\lim_{t\to\infty}\int \frac{d{\bf r}}{V}f({\bf q}(t, {\bf r})=
\int \frac{d{\bf r}}{V}f({\bf r})-\int_0^{\infty}\langle
\omega(0)f(t)\rangle dt\ .\label{finlim1}
\end{eqnarray}
If we choose the initial state for (\ref{basic}) as a constant,
$n_0({\bf r})\equiv n(t=0, {\bf r})=1/V$, then (\ref{finlim1}) can
be written as follows:
\begin{eqnarray}&&
\lim_{t\to\infty}\int \frac{d{\bf r}}{V}f({\bf r})n(t, {\bf r})=
\int \frac{d{\bf r}}{V}f({\bf r})-\int_0^{\infty}\langle
\omega(0)f(t)\rangle dt.
\end{eqnarray}
This suggests that the finiteness of temporal correlations i.e.
the existence of the integrals $\int_0^{\infty}\langle
\omega(0)f(t)\rangle dt$ for continuous $f$  is equivalent to the
existence of the limiting non-equilibrium state characterized by
the probability measure $\mu_{{\rm lim}}= \lim_{t\to\infty}n({\bf
r},t)$. Note that the difference of the non-equilibrium state
measure $\mu_{{\rm lim}}$ and equilibrium measure $E$ satisfies
\begin{eqnarray}&&
\mu_{{\rm lim}}(f)-E(f)=-\int_0^{\infty}\langle
\omega(0)f(t)\rangle dt,\ \ \mu_{{\rm lim}}(f)=\int fd\mu_{{\rm
lim}}, \ \ E(f)=\int \frac{d{\bf r}}{V}f({\bf r}).
\end{eqnarray}
Let us now consider Equation (\ref{cor1}). It involves a weaker
limit \cite{Ruelle}
$$\mu_{{\rm av}}=\lim_{t\to\infty}{1\over t}\int_0^tn({\bf r},t')\,dt'\ .$$
The existence of this limit is equivalent to the convergence of
the subtracted correlation integral
\begin{eqnarray}&&
\mu_{{\rm av}}(f)-E(f)=-\lim_{t\to\infty}\left[ \int_0^t \langle
\omega(0)f(t')\rangle dt'-\frac{1}{t}\int_0^t t' \langle
\omega(0)f(t')\rangle dt'\right]. \label{avera}\end{eqnarray} We
have $\mu_{{\rm av}}(f)=\mu_{{\rm lim}}(f)$ where $\int_0^{\infty}
\langle \omega(0)f(t)\rangle dt$ exists.

A limit of the type used in $\mu_{{\rm av}}$ appears where one
considers the sum of Lyapunov exponents $\sum\lambda_i({\bf r})$
(for the discussion of Lyapunov exponents $\lambda_i$ see
\cite{Oseledec,OGS}). That sum determines the growth rate of an
infinitesimal volume initially located at ${\bf r}$
\begin{eqnarray}&&
\sum \lambda_i({\bf r})=\lim_{t\to\infty}\frac{\ln \det W(t, {\bf
r})}{t}= \lim_{t\to\infty}\frac{1}{t}\int_0^t \omega({\bf q}(t',
{\bf r})dt'.
\end{eqnarray}
We observe that $\sum \lambda_i$ is represented as a time-average
of a function on the phase space. This is a unique combination of
$\lambda_i$ representable in such a form. Using Equation
(\ref{cor1}) we have
\begin{eqnarray}&&
\int \frac{d{\bf r}}{V}\sum \lambda_i({\bf
r})=-\lim_{t\to\infty}\left[ \int_0^t \langle
\omega(0)\omega(t')\rangle dt'-\frac{1}{t}\int_0^t t' \langle
\omega(0)\omega(t')\rangle dt'\right]=-\int_0^{\infty} \langle
\omega(0)\omega(t)\rangle dt,
\end{eqnarray}
where the last equality holds provided the integral exists. We
used $E(\omega)=0$ assuming that the integral over the boundary
vanishes. The above formula holds for systems whose stationary
measure is arbitrarily far from the equilibrium one and
nevertheless it has a remarkable resemblance to the Green-Kubo
formula holding near equilibrium. Note that it suggests that $\int
d{\bf r}\sum \lambda_i\leq 0$ always. This will be proved in the
next section. Here we note that $\int d{\bf r}\sum \lambda_i<0$
signifies that $\mu$ is singular. Therefore, the criterium of
singularity of the non-equilibrium measure is $\int_0^{\infty}
\langle \omega(0)\omega(t)\rangle dt>0$.

The above relations are simplified for systems satisfying the SRB
theorem that guarantees the equality between temporal average and
average with respect to the limiting measure for any continuous
$f$:
\begin{eqnarray}&&
\lim_{t\to\infty}\frac{1}{t}\int_0^t f({\bf q}(t', {\bf
r}))dt'=\int fd\mu_{{\rm SRB}}.
\end{eqnarray}
The above limit holds for almost all ${\bf r}$ in the sense of the
usual volume in Euclidean space and $\mu_{{\rm SRB}}$ is called
the SRB measure \cite{Sinai,BR75}. If the theorem holds we have
$$ \lim_{t\to\infty}\frac{1}{t}\int_0^t \langle g(0)f(t')\rangle
dt'=E(g)\mu(f),\ \ \lim_{t\to\infty}\frac{1}{t}\int_0^t \langle
\omega(0) f(t')\rangle dt'=0\ .$$ Note that $\mu_{{\rm SRB}}$
satisfies the relation (\ref{avera}) satisfied by $\mu_{{\rm
av}}$. For SRB-theorem systems $\sum\lambda_i({\bf r})$ is
constant almost everywhere and we have
\begin{eqnarray}&&
\sum \lambda_i=-\int_0^{\infty} \langle \omega(0)\omega(t)\rangle
dt.
\end{eqnarray}

We now apply the relations of this section for the investigation
of the evolution of the entropy which  possesses some unique
properties.

\section{Entropy}

The evolution of the entropy is closely related to the evolution
of infinitesimal volumes in the phase space. Indeed, let us
associate with each fluid element initially located at ${\bf r}$
the density of its Gibbs entropy (\ref{entropy1}):
\begin{eqnarray}&&
s({\bf r}, t)=-\ln n({\bf q}(t, {\bf r}), t),\ \ S(t)=\int d{\bf
r}\, n_0({\bf r})s({\bf r}, t). \label{baza1}\end{eqnarray} where
$n_0({\bf r})=n({\bf r}, t=0)$. The same quantity $\omega({\bf
q}(t, {\bf r}))$ determines the time derivative of both the
entropy density and the logarithm of the ratio of infinitesimal
volumes $v(t)/v(0)$: \begin{eqnarray}&& {\partial s\over\partial
t}=\omega({\bf q}(t, {\bf r}))={\partial \ln \det W(t, {\bf
r})\over\partial t}.
\end{eqnarray}
We conclude that the entropy associated with each fluid element
satisfies
\begin{eqnarray}&&
s({\bf r}, t)-s({\bf r}, 0)=\int_0^t \omega({\bf q}(t', {\bf
r}))\,dt',\ \ \lim_{t\to\infty}\frac{s({\bf r}, t)-s({\bf r},
0)}{t}=\sum\lambda_i. \label{infdens}\end{eqnarray} This property
of evolution of $s({\bf r}, t)$ holds for almost all ${\bf r}$ in
the limit $t\to\infty$. At finite times however one can observe
arbitrarily large deviations of $[s({\bf r}, t)-s({\bf r}, 0)]/t$
from $\sum\lambda_i$. These deviations are the subject of the
Evans-Searles formula which simple derivation within our formalism
we now present. Our approach here is based on the trivial
observation that ${\bf q}(t, {\bf q}(-t, {\bf r}))={\bf r}$.
Differentiating that we derive a useful identity \cite{BFF99}
\begin{eqnarray}&&
\frac{\partial q_i(t, {\bf r})}{\partial r_j}_{{\bf r}= {\bf
q}(-t, {\bf r})}=\left(\frac{\partial q_i(-t, {\bf r})} {\partial
r_j}\right)^{-1}.
\end{eqnarray}
We can now relate the volume fraction of the trajectories that
during time $t$ change their entropy density by some $\Delta$ to
the volume fraction of the trajectories with the change $-\Delta $
under the time-reversed evolution. We have
\begin{eqnarray}&&
\int \frac{d{\bf r}}{V}\,\delta\left(-\ln \det
\partial_jq_i(-t, {\bf r})+\Delta \right)\det
\partial_jq_i(-t, {\bf r})=\int \frac{d{\bf r}}{V}\,\delta
\left(-\ln \det \partial_jq_i(t, {\bf r})|_{{\bf r}= {\bf q}(-t,
{\bf r})}-\Delta \right)\det \partial_jq_i(-t, {\bf r}) \nonumber
\\&&
=\int \frac{d{\bf r}}{V}\,\delta \left(-\ln \det
\partial_jq_i(t, {\bf r})-\Delta \right)
\end{eqnarray}
It follows that the measures of $\Delta s(t, {\bf r})\equiv s(0,
{\bf r})-s(t, {\bf r})$ obey
\begin{eqnarray}&&
\frac{\int d{\bf r}\,\delta\left(\Delta s(t, {\bf r})-\Delta
\right)} {\int d{\bf r}\,\delta\left(\Delta s(-t, {\bf r})+\Delta
\right)}=e^\Delta\ .
%,\ \ \epsilon(t, {\bf r})=\frac{s(0, {\bf r})-s(t, {\bf r})} {|t|}.
\label{gen}\end{eqnarray} In the case of time-reversible dynamics,
$\int {d{\bf r}}\,\delta\left(\Delta s(-t, {\bf r})+\Delta
\right)= \int {d{\bf r}}\,\delta\left(\Delta s(t, {\bf r})+\Delta
\right)$ and
\begin{eqnarray}&&
\frac{\int d{\bf r}\,\delta\left(\Delta s(t, {\bf r})-\Delta
\right)} {\int d{\bf r}\,\delta\left(\Delta s(t, {\bf r})+\Delta
\right)}= e^\Delta\ , \label{gen1}\end{eqnarray} which says that,
starting from equilibrium, the probability to have the entropy
change $\Delta$ is by factor $\exp(\Delta )$ larger than the
probability to have the entropy change $-\Delta$. That relation
was first established by Evans and Searles \cite{ES}. The
probability is defined here with respect to Lebesgue measure (the
volume fraction occupied by respective trajectories) and the
relation is valid for any time. Recasting the above equation in
the form
\begin{eqnarray}&&
\frac{\int d{\bf r}\delta\left(\epsilon(t, {\bf r})-a\right)}
{\int d{\bf r}\delta\left(\epsilon(t, {\bf r})+a\right)}=
\exp\left[a t\right]\ ,\ \ \epsilon(t, {\bf r})=\frac{\Delta s(t,
{\bf r})} {|t|} \label{gen3}\end{eqnarray} one can compare it to
the Gallavotti-Cohen formula \cite{GC}:
\begin{eqnarray}&&
\frac{\int d\mu\delta\left(\epsilon(t, {\bf r})-a\right)} {\int
d\mu\delta\left(\epsilon(t, {\bf r})+a\right)}= \exp\left[a
t\right]\ , \label{gen4}\end{eqnarray} that holds at large $t$.
Probability  in ({\ref{gen4}) is defined with respect to the SRB
measure. Comparison between (\ref{gen3}) and (\ref{gen4}) is
discussed in \cite{CG1}.

Let us now consider more closely the density of the entropy
production rate $\partial_t s(t, {\bf r})=\omega({\bf q}(t, {\bf
r}))$. According to the general formula (\ref{deriv}) we have
\begin{eqnarray}&&
\partial_t^2 s(t, {\bf r})=-\omega({\bf r})\omega({\bf q}(t, {\bf r}))
+\nabla_r\cdot [{\bf v}({\bf r})\omega({\bf q}(t, {\bf r}))].
\label{yo}\end{eqnarray} Both $\omega({\bf q}(t, {\bf r}))$ and
its time derivative can have arbitrary sign. However, the above
formula for $\partial_t^2 s(t, {\bf r})$ suggests that the global
quantities like space integrals can exhibit some general
properties. The relation (\ref{infdens}) for the entropy density
implies an analogous relation for the entropy
\begin{eqnarray}&&
S(t)-S(0)=\int_0^tdt' \int d{\bf r}n_0({\bf r}) \omega({\bf q}(t',
{\bf r})),\ \ \lim_{t\to\infty}\frac{S(t)-S(0)}{t}=\sum\lambda_i.
\label{infdens1}\end{eqnarray} Note that for systems with ${\bf
r}$-independent $\sum\lambda_i({\bf r})$ the last expression
equals $\sum\lambda_i$. Generally, considering the case $n_0({\bf
r})=const$ we find
\begin{eqnarray}&&
\sum\lambda_i\leq 0\,,\label{pos}
\end{eqnarray}
which is expected from the Equation (\ref{cor}). Indeed since
among all the densities having a given normalization the maximal
 $S$ corresponds to a constant density (see Appendix
\ref{static}) and the evolution conserves normalization, we have
$S(t)-S(0)\leq 0$ with equality holding only if the flow brings
the density back to the constant at some time $t$. The relation
(\ref{pos}) was demonstrated before \cite{Ruelle97,BFF99}. Another
interesting quantity is the entropy production rate that can be
written as
\begin{eqnarray}&&
\Gamma(t)\equiv \frac{dS}{dt}=\int n_0({\bf r}) \omega({\bf q}(t|
{\bf r}))d{\bf r}=\int d{\bf r}n({\bf r}, t)\omega({\bf r})
\label{e1}
\end{eqnarray}
For an arbitrary $n_0({\bf r})$, the entropy production rate can
have any sign at small $t$. The most interesting case is the case
of $n_0({\bf r}) =const$ where the evolution of $n({\bf r}, t)$ is
that from equilibrium to the non-equilibrium steady state
characterized by the SRB measure. In this case we have
\begin{eqnarray}&&
\Gamma(t)=\int\frac{d{\bf r}}{V}\omega({\bf q}(t| {\bf r})).
\end{eqnarray}
Clearly $\Gamma(0)=0$. Under the condition of sufficiently fast
temporal decay of correlations (existence of $\int_0^{\infty}t
dt\langle \omega(0) \omega(t)\rangle$) it follows from the
analysis of the Sect.~\ref{sec:generalissimus} that
\begin{eqnarray}&&
\lim_{t\to \infty}\Gamma(t)=\sum\lambda_i\leq 0.
\end{eqnarray}
The case $\sum \lambda_i=0$ is the case where the steady
non-equilibrium state $\mu$ does not exchange entropy with the
environment. We wish to consider $\sum \lambda_i<0$. Then
$\Gamma(t=\infty)<0$. We also should have $\Gamma(t)<0$ at small
$t$ since at these times $n(t, {\bf r})$ goes away from the
constant giving the maximum to $S$. It is natural then to ask what
flows correspond to $\Gamma(t)$ being a monotonic decreasing
function of $t$ reaching its minimum in the non-equilibrium steady
state. This property cannot be general: if the flow brings back
$n(t, {\bf r})$ close to its original constant value (not exactly
the same to have $\int d{\bf r}\sum \lambda_i<0$) then $\int_0^t
\Gamma(t')dt'\approx 0$ and $\Gamma(t)$ changes sign.
 To
understand better the evolution of $\Gamma(t)$ consider
\begin{eqnarray}&&
\!\!\!\!\!\!\!\!\frac{d\Gamma}{dt}=\int d{\bf r} n_0({\bf
r})[({\bf v}({\bf x})\cdot\nabla_x) \omega({\bf x})]_{{\bf x}=
{\bf q}(t| {\bf r})}=- \int d{\bf r}n_0({\bf r}) \omega({\bf r})
\omega({\bf q}(t| {\bf r}))+\int d{\bf r}n_0({\bf
r})\nabla_r\cdot[{\bf v}({\bf r}) \omega({\bf q}(t| {\bf r}))]\,,
\end{eqnarray}
where we used the formula (\ref{yo}). We observe that for ${\bf
v}$ such that ${\bf v}\cdot\nabla\omega<0$ everywhere we have
$\dot{\Gamma}(t)<0$ and the conjecture on the monotonic decrease
of $\Gamma$ holds. Note that the opposite case ${\bf
v}\cdot\nabla\omega>0$ would violate the possibility to integrate
by parts $\int {\bf v}\cdot\nabla\omega=-\int \omega^2<0$.

We now concentrate on $n_0({\bf r})=1/V$ case where
\begin{eqnarray}&&
\frac{d\Gamma}{dt}=-\langle\omega(0)\omega(t)\rangle, \ \
\Gamma(t)=-\int_0^t \langle\omega(0)\omega(t')\rangle dt'
\end{eqnarray}
Clearly at small $t$ we have $\dot{\Gamma}(t)<0$. The above
relation allows us to introduce the criterium of the
generalization of the principle of maximal generation of entropy
that holds near equilibrium. If the system possesses positive
everywhere correlation function $\langle\omega(0)\omega(t)\rangle$
(that implies, in particular $\sum \lambda_i<0$) then on its way
from equilibrium to the non-equilibrium steady state the system
gives more and more entropy to the environment with the rate that
reaches maximum in the steady state.

Let us summarize our results on the entropy production rate
$\Gamma(t)$. The first question is the sign-definiteness of
$\Gamma(t)$. We argued that even for $n_0({\bf r})=1/V$ such
assertion cannot be made generally. It seems that the correct
language to consider this question and the question of the
generalization of the maximum entropy production principle to the
strongly non-equilibrium situations is to consider
$\dot{\Gamma}(t)$. We have shown that the latter is equal to minus
the value of an auto-correlation function at time difference $t$.

\section{Generalization for a time-dependent velocity with stationary
statistics}

We generalize the above results to the case of time-dependent
velocity ${\bf v}(t, {\bf r})$. The phase-space trajectories ${\bf
q}(t | t', {\bf r})$ depend now on both the initial time $t'$ and
the final time $t$ and are characterized by
\begin{eqnarray}&&
\frac{\partial {\bf q}(t | t', {\bf r})}{\partial t}={\bf v} ({\bf
q}(t | t', {\bf r}), t),\ \ {\bf q}(t' | t', {\bf r})={\bf r}\ .
\end{eqnarray}
We introduce $$\ \ W_{ij}(t | t', {\bf r})=\frac{\partial q_i(t |
t', {\bf r})}{\partial r_j}\ .$$ To establish the generalization
of (\ref{gen}) we use the same proof replacing the function ${\bf
q}(-t, {\bf r})$ by ${\bf q}(0 | t, {\bf r})$. For example we have
\cite{BFF99}
\begin{eqnarray}&&
\frac{\partial q_i(t| 0, {\bf r})}{\partial r_j}_{{\bf r}= {\bf
q}(0| t, {\bf r})}=\left(\frac{\partial q_i(0 | t, {\bf r})}
{\partial r_j}\right)^{-1}.
\end{eqnarray}
We find the identity
\begin{eqnarray}&&
\frac{\int  d{\bf r}\,\delta \left(-\ln \det
\partial_jq_i(t| 0, {\bf r})/t-a\right)}{\int  d{\bf r}\,\delta \left(-\ln \det
\partial_jq_i(0 | t, {\bf r})/t+a\right)}=\exp[at]
\end{eqnarray}
that holds for any ${\bf v}(t, {\bf r})$. If ${\bf v}(t, {\bf r})$
has stationary statistics, with the average defined by the spatial
integration, then one can replace $q_i(0 | t, {\bf r})$ by
$q_i(-t| 0, {\bf r})$ in the denominator of the above formula and
the formula (\ref{gen}) results. Moreover if the statistics is
time-reversible we arrive at the equation
\begin{eqnarray}&&
\frac{\int d{\bf r}\,\delta\left(\epsilon(t, {\bf r})-a\right)}
{\int d{\bf r}\,\delta\left(\epsilon(t, {\bf r})+a\right)}=
\exp\left[a t\right]\ , \label{gen2}\end{eqnarray} completing the
generalization.

Next we consider the entropy production rate for the case
$n_0({\bf r})=1/V$
\begin{eqnarray}&& \Gamma(t)=\int \omega(t, {\bf q}(t,
{\bf r}))\frac{d{\bf r}}{V}. \label{e5} \end{eqnarray} Direct time
differentiation of the above equation is not very useful in this
case: we have no control over $\partial_t \omega$. The idea is to
manipulate the integral recasting it in the form more suitable for
the time-differentiation. Changing variables ${\bf y}={\bf q}(0|
t, {\bf r})$ we find
\begin{eqnarray}&&
\Gamma(t)=\int \omega(t, {\bf y}) \exp\left[-\int_0^t \omega({\bf
q} (t'| t, {\bf y}), t')dt'\right]\frac{d{\bf y}}{V}\ .\label{one}
\end{eqnarray} We now assume stationarity that is we assume that
the above average is equal to
\begin{eqnarray}&& \Gamma(t)=\int
\omega({\bf y}, 0)\exp\left[-\int_{-t}^0 \omega({\bf q} (t'| {\bf
y}), t')dt'\right]\frac{d{\bf y}}{V}\ .\label{two} \end{eqnarray}
This allows us to avoid the appearance of $\partial_t\omega$ in
$\dot{\Gamma}$. We have
\begin{eqnarray}&& \frac{d\Gamma}{dt}=-\int
\omega({\bf y}, 0)\omega({\bf q}(-t| {\bf y}), -t)
\exp\left[-\int_{-t}^0 s({\bf q} (t'| {\bf y}), t')dt'\right]
\frac{d{\bf y}}{V}=-\langle\omega(0)\omega(t)\rangle,
\end{eqnarray}
where the last equality is obtained by getting back in the
integral to the variable ${\bf r}={\bf q}(t| 0, {\bf y})$. The
corresponding expression for the sum of Lyapunov exponents
follows. This completes the generalization of the main results
derived for the steady velocity field to the case of the
time-dependent velocity with stationary statistics. If integrating
over space does not guarantee stationarity, i.e. coincidence of
(\ref{one},
\ref{two}), and one needs to add average over velocity
ensemble,
 then (\ref{d1s}) and (\ref{gen2}) are also valid after such ensemble average.

G.F. is grateful to D. Ruelle for helpful explanations. We thank
K. Gawedzki for reading the manuscript and providing us with
useful remarks. The work is supported by the Minerva foundation.
%\section{Summary}

%\begin{thebibliography}

\appendix

\section{Properties of the Gibbs entropy}
\label{static}

We make a general remark on the entropy properties which are not
related to the dynamics. We use the fact that $\ln x-1+1/x\geq 0$
(here $x>0$) with the equality reached only at $x=1$. It follows
that for any two normalized functions $n$ and $n'$ we have
\begin{eqnarray}&&
-\int n\ln\left(\frac{n}{n'}\right)d{\bf r}\leq 0,
\end{eqnarray}
with the equality reached iff $n=n'$. The above inequality says
that entropy considered as expectation value over some probability
distribution $n$ of minus logarithm of arbitrary other
distribution $n'$ is minimal where $n'=n$. An interesting
conclusion follows for a compact space with volume $V$. We have
for any distribution $n$
\begin{eqnarray}&&
\int n\ln \left(n V\right)d{\bf r}\geq 0,
\label{forward}\end{eqnarray} with the equality reached iff $n$ is
uniform. A compact space is analogous to the probability space
consisting of finite number of events in that one can define the
entropy that is bounded from above by the maximal value reached at
the uniform distribution. We also have that
\begin{eqnarray}&&
\int \frac{d{\bf r}}{V}\ln \left(n V\right)\leq 0.
\label{backward}\end{eqnarray} In particular, using the above
equations for $n=n({\bf r}, t)$ we find from Eq. (\ref{forward})
that the sum of forward in time Lyapunov exponents is non-positive
while the last equation implies the same conclusion about the sum
of backward in time Lyapunov exponents.

\end{document}